\documentclass[twocolumn]{aastex63}

\submitjournal{ApJ}

\begin{document}
\title{Delayed Detonation Thermonuclear Supernovae With An Extended Dark Matter Component}

\author{Ho-Sang Chan}
\affiliation{Department of Physics and Institute of Theoretical Physics, The Chinese University of Hong Kong, Shatin, N.T., Hong Kong S.A.R., China}
\author{Ming-chung Chu}
\affiliation{Department of Physics and Institute of Theoretical Physics, The Chinese University of Hong Kong, Shatin, N.T., Hong Kong S.A.R., China}
\author{Shing-Chi Leung}
\affiliation{TAPIR, Walter Burke Institute for Theoretical Physics, Mailcode 350-17, Caltech, Pasadena, CA 91125, USA}
\author{Lap-Ming Lin}
\affiliation{Department of Physics and Institute of Theoretical Physics, The Chinese University of Hong Kong, Shatin, N.T., Hong Kong S.A.R., China}

\begin{abstract}
We present simulations of thermonuclear supernovae admixed with an extended component of fermionic cold dark matter. We consider the explosion of a Chandrasekhar-mass white dwarf using the deflagration model with deflagration-detonation transition with spherical symmetry. The dark matter component is comparable in size with that of the normal matter, and so the system is described by two-fluid, one-dimensional Eulerian hydrodynamics. The explosion leaves all the dark matter trapped as a remnant compact dark star in all of our considered models. The presence of dark matter lengthens the deflagration phase to produce more thermo-neutrinos and similar amounts of iron-group elements compared to those of ordinary explosions with no dark matter admixture. The dark matter admixed models produce dimmer and broader light curves, which challenge the role of thermonuclear supernovae as standard candles in cosmic distance measurement. Our results also suggest a formation path of dark compact objects which mimic sub-solar-mass black holes as dark gravitational sources, through near-solar-mass dark matter admixed thermonuclear supernovae.
\end{abstract}


\section{Introduction} \label{sec:intro}
\subsection{Dark Matter Astronomy} \label{subsec:dmastronomy}
Dark matter (DM) is believed to be the dominant component of galaxy clusters \citep{Clowe_2006}, large scale structures of the universe \citep{1985ApJ...292..371D}, as well as a host of other astrophysical objects \citep{Oppenheimer698}. Yet the searches for DM particles have not yielded convincing candidates. The Large Hadron Collider searches for DM signals are still on-going. The possibly positive signals from the Xenon1T DM detector \citep{aprile2020observation} may provide hints to the existence of DM. Nevertheless, astrophysical searches for DM will remain an important channel to complement terrestrial experiments. 

\subsection{DM Admixed Stellar Objects} \label{subsec:dmstellar}
Given that DM is a major component in the universe, stellar objects comprising normal matter (NM) and DM may form. Study of self-interacting and non-self-interacting DM-admixed neutron stars suggest that compact stars can be a probe to DM \citep{PhysRevD.81.123521}. Later studies on the equilibrium structures of non-self-interacting DM-admixed neutron stars provided hints to search for DM from neutron star diversities \citep{PhysRevD.84.107301, PhysRevD.97.123007}. On-going studies focus on different perspectives of compact stars including bosonic DM stars \citep{Eby_2016}, formation of a neutron star through accretion-induced collapse of a DM-admixed white dwarf \citep{Leung_2019,Zha_2019}, thermalization of the white dwarf (WD) core to produce thermonuclear runaway \citep{PhysRevLett.115.141301,PhysRevD.100.043020,PhysRevD.100.035008,steigerwald2019dark}, and  point-mass DM-admixed in thermonuclear supernovae \citep{Leung_2015}. Recent studies have been made on DM-admixed pulsars \citep{article}, effects of DM admixture on star formation \citep{Arun_2019}, bosonic DM-admixed neutron stars with relativistic Bose-Einstein Condensation (Lee et al., in prep) and tidal Love numbers of DM-admixed neutron stars (Leung et al., in prep). DM-admixed astrophysical objects have become a new window to search for astrophysical DM. 

\subsection{DM Admixed Thermonuclear Supernovae} \label{subsec:dmsupernova}
\citet{PhysRevD.100.043020}, \citet{PhysRevLett.115.141301} and \citet{steigerwald2019dark} pointed out that accretion of DM to a WD could lead to kinetic energy transfer from DM to NM through scattering, which helps to thermalize the NM core towards the temperature of thermonuclear explosion. On the other hand, \citet{PhysRevD.87.123506} showed that the admixture of non-interacting DM could reduce the Chandrashekhar limit of WDs, by increasing the NM central density so that the thermonuclear flame is generated already when the mass of the WD is sub-Chandrasekhar. \\

\citet{Leung_2015} have demonstrated that DM-admixed thermonuclear supernovae produce sub-luminous light curves, by assuming a heavy fermionic DM particle mass of around $1$ GeV. The resulting DM core is small enough to be approximated as a stationary point gravity source at the WD center. Recent constraints on sub-GeV DM derived from lab experiments \citep{cirelli2020integral} and neutron star masses \citep{ivanytskyi2019neutron} allow for sub-GeV DM particle models. Many extensions of the Standard Model of particle physics also give possibilities for sub-GeV DM particles \citep{PhysRevD.100.075028}. We are therefore motivated to study the effects of admixture of sub-GeV DM particles on thermonuclear supernovae. As we shall see, sub-GeV DM particles create an extended DM component which could not be treated as a point mass, and so we extend the explosion hydrodynamics using a two-fluid formalism.

\subsection{The Deflagration-Detonation Transition As Supernova Explosion Model} \label{subsec:ddtsupernova}
There is a high degree of homogeneity in the spectra of thermonuclear supernovae \citep[see, for example, the review of][]{doi:10.1146/annurev.astro.38.1.191}, which suggests that WDs should be their common progenitors \citep{hillebr2013understanding}. Early thermonuclear supernova simulations investigated pure detonation driven explosions \citep{ArnettW.David1969Apmo} and pure deflagration explosions \citep{NomotoKen&apos;ichi1976Cdsa}. The former found that iron-group elements were over-produced, while the latter found that pure laminar deflagration was too slow / sub-sonic to generate a healthy explosion. Results from either pure detonation or pure laminar deflagration models do not fully reconcile with the observed features in a typical thermonuclear supernova. Moreover, it was argued that pure detonation will be quenched at high density, casting doubt on the possibilities of prompt detonation \citep{KRIMINSKI1998363}. \\

On the other hand, deflagration is subject to various hydrodynamical instabilities \citep{hillebr2013understanding}. Turbulence induce convolutions on the flame surface, which allow the flame to propagate in an effective speed faster than its laminar flame speed \citep{1984ApJ...286..644N,BranchDavid2017Se}. Early studies of turbulent deflagration using the mixing-length theory by \citet{NomotoKen&apos;ichi1976Cdsa} and \citet{1984ApJ...286..644N} (The W7 model), which assume fast deflagration around $20$ -- $30$ \% of local sound speed, can produce healthy explosions, but they also result in over-production of $^{58}$Ni, in conflict with constraints from galactic chemical evolution \citep{Iwamoto_1999}\footnote{However, \citet{Iwamoto_1999} also argued that the electron capture rates they employed are too high and may be an another reason for over-production of $^{58}$Ni.}. Multi-dimensional studies on pure turbulent deflagration also under-produce $^{56}$Ni and explosion energy \citep{reinecke1998thermonuclear, calder2004type}, with most of the unburned material remaining at the center which should only exist at the outer ejecta \citep{Gamezo77}. Although later studies showed that good agreement with observations in terms of nickel mass and ejecta velocity could be obtained \citep{levelset3d, ropke_2005}, another major shortcoming of the pure turbulent deflagration model is the underproduction of silicon-group elements in the outer layers of the ejecta \citep{ropke_2007}. \\

A transition to detonation from deflagration is observed in several terrestrial experiments with closed boundaries \citep{BranchDavid2017Se}. It is therefore natural to hypothesize that transitions to detonation also happen in open boundaries such as WDs. \citet{1991A&A...245..114K} proposed the deflagration-detonation transition (DDT) model which shows comparable amount of iron-group and silicon-group elements to observed thermonuclear supernovae. DDT models are also shown to fit observed supernova light curves in V and R bands reasonably well \citep{1995ApJ...444..831H}. To trigger DDT, the Zel’dovich gradient mechanism \citep{ZeldovichYa.B1972Otoo} is necessary, where local eddy motion near the flame front flattens the temperature gradient \citep{Niemeyer_1999, BranchDavid2017Se}. Whether turbulence can attain the required temperature gradient robustly is a concern \citep{Niemeyer_1999, Lisewski_2000, Woosley_2007}. In principle, the turbulence motion can mix fuel and ash to obtain the required pre-conditioned field \citep{hillebr2013understanding}. However, numerical simulations show that the implied velocity fluctuation is only marginally sufficient for triggering the DDT \citep{Ropke_2007_DDT}. This casts doubts on whether DDT could robustly occur in thermonuclear explosions.\\

The feasibility of turbulent flame and DDT has led to more theoretical models, accompanied with the growing diversity of observational data, such as gravitational confined detonation model \citep{Plewa_2004} and pulsating reverse detonation models \citep{Bravo_2006}. 

\subsection{Motivation} \label{subsec:motivate}
In \citet{Leung_2015} the DM-admixed thermonuclear supernova is studied by considering the turbulent deflagration model without DDT. In this work, we extend the study of \citet{Leung_2015} by (1) focusing on the successful DDT model and (2) adding the dynamics of the DM component in the picture using fully non-linear two-fluid simulations. The plan of the paper is as follows: Section \ref{sec:method} describes the constructions of supernova progenitor models and the tools for simulating the explosions and post-processing methods. Section \ref{sec:results} summarizes the simulation results. Section \ref{sec:discussion} discusses the parameter dependence of our models and the implications of our results, and Section \ref{sec:conclusion} concludes our study.

\section{Methodology} \label{sec:method}
\subsection{Equilibrium Structure Of DM-admixed WDs} \label{subsec:dmawd}
We construct a series of DM-admixed WDs (DMWD) as supernova progenitors, using the Newtonian hydrostatic equilibrium equations \citep[see, for example,][]{SANDIN2009278, CIARCELLUTI201119, Leung_2019}:\footnote{Though the former two works are based on relativistic formulation, the equations of hydrostatic equilibrium can be obtained by taking the non-relativistic limit.}
\begin{equation}
	\frac{dp_{i}}{dr} = -\frac{G(m_{1}+m_{2})\rho_{i}}{r^{2}},
\end{equation}
\begin{equation}
	\frac{dm_{i}}{dr} = 4\pi r^{2}\rho_{i}.
\end{equation}
Here, the subscript $i = 1(2)$ denotes DM (NM) quantities, and $\rho$, $p$, $m$ and $r$ are the density, pressure, enclosed mass and radial coordinate respectively. We choose the ideal degenerate Fermi gas equation of state (EOS) assuming spin $\frac{1}{2}$ for DM particles \citep{TeukolskySaulA2008BHWD}. We vary the DM particle mass from $0.1$ GeV to $0.3$ GeV in $0.05$ GeV steps. For each DM particle mass, we vary the DM central density from $10^{8}$ g cm$^{-3}$ to $10^{10}$ g cm$^{-3}$. We will show how we obtain the supernova progenitors by analysing the series of models.

\subsection{Hydrodynamic Solver} \label{subsec:hydrosolver}
Since DM only affects the dynamics of NM by gravity, we take a first step in this study to identify the primary effects of DM on thermonuclear supernovae by considering spherically symmetric Newtonian two-fluid hydrodynamics in the Eulerian framework. Multi-dimensional two-fluid simulations in this scenario are computationally demanding, because during the late explosion phase, a large simulation box is required when the NM can extend to a size much larger than its initial value, while a fine enough grid resolution is needed to resolve the central DM component. In typical one-dimensional simulations Lagrangian formalism is often used. We do not consider this formalism because further interpolation of the local gravitational force by both matter becomes necessary when the fluid elements do not overlap with each other. The two-fluids may also have different masses and radii, which makes the definition of mass coordinate difficult. \\

Our supernova code is constructed based on a two-fluid hydrodynamics code developed by \citet{WongKaWing2011Ppow}. Unless otherwise noted, we use geometric units with $G=c=M_{\odot}=1$. Different from \citet{Leung_2015}, we adopt the finite-volume approach where the Euler equations in spherical coordinate are given in \citet{MIGNONE2014784}:
\begin{equation}
	\frac{d}{dt}\vec{U}_{i} = - \frac{1}{r^{2}}\frac{d}{dr}(r^{2}\vec{F}_{i}) + \vec{S}_{i} + \vec{G}_{i},
\end{equation}
with the state vector $\vec{U}$ and flux vector $\vec{F}$ given as
\begin{eqnarray}
\vec{U}_{i} = \left(
\begin{array}{c}
\rho_{i} \\
\rho_{i}v_{i} \\
E_{i} \\
\Psi_{i}
\end{array}\right),
\end{eqnarray}
\begin{eqnarray}\
\vec{F}_{i} = \left(
\begin{array}{c}
\rho_{i}v_{i} \\
\rho_{i}v_{i}^{2} + p_{i} \\
(E_{i} + p_{i})v_{i} \\
\rho_{i}\Psi_{i}
\end{array}\right).
\end{eqnarray}
Here $\epsilon_{i}$ is the internal energy, $v_{i}$ is the radial velocity and $E_{i} = \rho_{i}(\epsilon_{i} + \frac{1}{2}v_{i}^{2})$ is the total energy density. $\Psi_{i}$ is any passively scalar quantity. For example, $\Psi_{i}$ could be the isotope mass fractions $X(^{12}$C) of $^{12}$C. The geometric source term $\vec{S}$ and extra source term are given as
\begin{eqnarray}\
\vec{S}_{i} = \left(
\begin{array}{c}
0 \\
2p_{i}/r \\
0 \\
0
\end{array}\right),
\end{eqnarray}
\begin{eqnarray}\
\vec{G}_{i} = \left(
\begin{array}{c}
0 \\
-\rho_{i}\frac{d\phi}{dr} \\
-\rho_{i}v_{i}\frac{d\phi}{dr} \\
0
\end{array}\right),
\end{eqnarray}
where $\phi$ is the gravitational potential, governed by the two-fluid Poisson's equation:
\begin{equation}
	\nabla^{2} \phi = 4\pi(\rho_{1}+\rho_{2}).
\end{equation}
We adopt the accurate monotonicity-preserving scheme named as Monotonicity-Preserving 5th Order (MP5) scheme \citep{SURESH199783} to reconstruct primitive variables at the cell interfaces. We choose the Lax-Friedrich Riemann solver \citep{JIANG1996202} to compute interface numerical fluxes and to discretize the temporal evolution using the method of lines where the strong stability-preserving 3rd-order Runge-Kutta's method is implemented \citep{doi:10.1142/7498}.\\

We adopt a modified monopole solver \citep{Couch_2013} for the gravitational potential where the error of self-gravity could be reduced. We implement a moving-grid algorithm \citep{roepke2004following} to follow the explosion ejecta until the expansion becomes homologous. The moving-grid equation introduces a grid velocity $v_{f}$, and the flux and source vectors are modified as follows \citep{Leung_2018}:
\begin{equation}
	\vec{F}_{i} \rightarrow \vec{F}_{i} - \vec{U}_{i}v_{f},
\end{equation}
\begin{equation}
	\vec{G}_{2} \rightarrow \vec{G}_{2} - \vec{U}_{2}\frac{1}{r^{2}}\frac{d}{dr}(r^{2}\vec{v_{f}}).
\end{equation}
To properly capture all nucleosynthesis before the material reaches the boundaries where the moving-grid algorithm is triggered, the initial grid box size is set to be $2.7\times10^{9}$ cm, which is more than $10$ times of the typical size of the progenitor configurations.

\subsection{Simplified Nuclear Network} \label{subsec:network}
To reduce computational time, we implement a reduced nuclear network used in the work by \citet{Leung_2018} and \citet{Leung_2020}, which was developed based on the original work by \citet{Townsley_2007} and \citet{Calder_2007}. The nuclear network includes $7$ isotopes - $^{4}$He, $^{12}$C, $^{16}$O, $^{20}$Ne, $^{24}$Mg, $^{28}$Si, $^{56}$Ni. The nuclear network separates nuclear burning into 3 stages:
\begin{enumerate}
	\item{Combustion of $^{12}$C to $^{24}$Mg;}
	\item{Burning of $^{16}$O and $^{24}$Mg to $^{28}$Si;}
	\item{Nuclear statistical equilibrium (NSE).}
\end{enumerate}
When the flame first sweeps through the fuel, step 1 is triggered. The burning proceeds to step 2 which is also called the nuclear quasi statistical equilibrium (NQSE) when the time step $\delta t$ is larger than the quasi equilibrium time step $\tau_{\rm NQSE} = \exp(182.06/T_{9} - 46.064)$, where $T_{9}$ is the temperature in $10^{9}$ K. Finally the burning proceeds to step 3 if the time step is larger than the NSE time scale $\tau_{\rm NSE} = \exp(196.02/T_{9} - 41.645)$. For incomplete burning ($\delta t < \tau_{\rm NQSE}$ or $\tau_{\rm NSE}$) we will use linear interpolation.

\subsection{Flame Capturing And Delayed Detonation} \label{subsec:levelset}
To capture the flame width which is $\sim 10^{-4}$ cm \citep{BranchDavid2017Se} and is propagating at a subsonic speed, we use a flame capturing scheme to follow the flame propagation. We adopt the level-set method \citep{SethianJamesAlbert1999Lsma} which has been used by several authors before \citep{reinecke1998new,levelset3d,refine}. To initiate a deflagration, we plant a level-set enclosing $\sim 0.02$ $M_{\odot}$ of NM for all models. The spherical symmetry does not allow us to compute turbulence production explicitly. We follow the method in the literature \citep{1984ApJ...286..644N,1995ApJ...444..831H,Woosley_1997,ddt_2012,ddt_2015} to parameterize the deflagration speed in terms of the local sound speed $c_{s}$. The typical value would be a fraction, possibly few percent of sound speed. After various numerical experiments of pure NM explosions, we choose the fraction to be $0.06$ so that the $^{56}$Ni produced is around $0.5$ -- $0.6$ $M_{\odot}$. The DDT is triggered once the flame front reaches a density $1.7\times10^{7}$ g cm$^{-3}$ \citep{Iwamoto_1999}. During the onset of DDT, the level-set speed is immediately raised to the detonation speed. The detonation for density lower than $\sim 2\times 10^{7}$ g cm$^{-3}$ is a CJ detonation for which the speed is the speed of sound, while the detonation for density higher than $\sim 2\times 10^{7}$ g cm$^{-3}$ is a pathological one. The latter is solved using the method by \citet{10.1046/j.1365-8711.1999.03023.x} and \citet{Leung_2020}.

\subsection{Supernova Observables} \label{subsec:observables}
Recent studies on thermonuclear supernovae show that neutrino production could be a key to distinguish explosion models \citep{PhysRevD.94.025026,PhysRevD.95.043006}.  It is thus likely that the neutrino signal also plays an important role in DM-admixed models. We use the open-source neutrino emission subroutine\footnote{http://cococubed.asu.edu/} which calculates pair, photo-, bremsstrahlung, and recombination neutrinos with formulae derived in \citet{1996ApJS..102..411I}. The plasmon neutrinos are computed using the method by \citet{10.1111/j.1365-2966.2007.12342.x}. To compute the differential production rate for high energy neutrinos ($1$ -- $5$ MeV), we adopt the method by \citet{PhysRevD.74.043006} to calculate the pair neutrino spectrum. In addition, the plasma neutrino spectrum is calculated using the method by \citet{OdrzywoekA2007Ps}. These methods have also been applied in our previous works \citep{Leung_2015_code, Leung_2020_PP}. The differential production rate is integrated to obtain the total number of thermo-neutrinos produced. \\

\citet{roepke2004following} found that the ejecta approach homologous expansion in between $\sim$5 -- 10 s. Furthermore, most exothermic nuclear reactions end after $\sim$ 1 s, where the isotope mass fraction distribution does not change much between $5$ s and $10$ s. Moreover, our numerical models with different termination times show that mass and energy conservation could become worse if we increase the simulation time due to significant increase in grid size. At $5$ s after the initial runaway has started, we terminate the Eulerian hydrodynamics simulation.  The resulting density, temperature, velocity and isotope  profiles of our simulations were mapped from our one-dimensional Eulerian form into one-dimensional Lagrangian form, which are then input to the SNEC code \citep{Morozova_2015} to calculate the light curve and dynamics during the post-explosion phase. 

\section{Results} \label{sec:results}
\subsection{Equilibrium Structure Of DMWD} \label{subsec:structure}
\begin{figure}[ht!]
	\centering
	\includegraphics[width=1.0\linewidth]{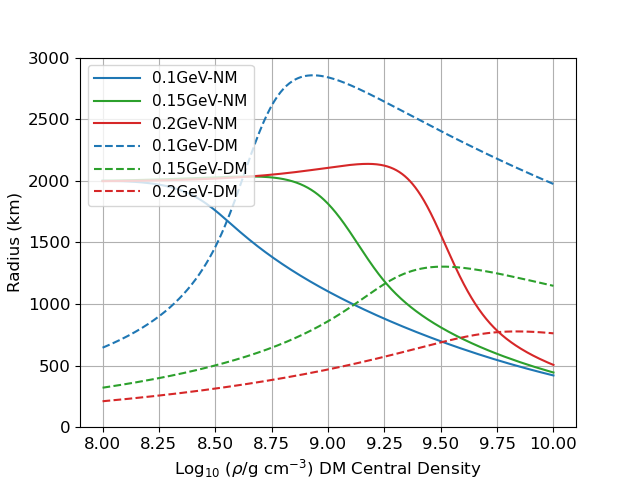}
	\caption{Radii of different components against DM central density for DM particle masses of 0.1 (blue), 0.15 (green) and 0.2 GeV (red). The solid (dashed) lines are for the NM (DM) component. The NM central density is kept at $3\times10^{9}$ g cm$^{-3}$. \label{fig:radrhoc}}
\end{figure}
\begin{figure}[ht!]
	\centering
	\includegraphics[width=1.0\linewidth]{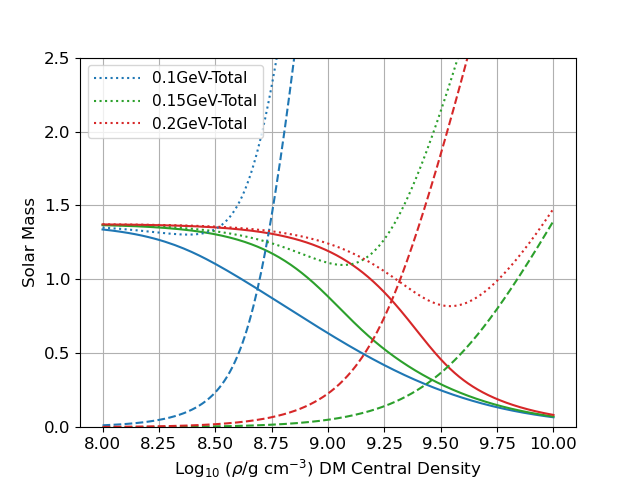}
	\caption{Same as Figure \ref{fig:radrhoc}, but for the masses. The dotted lines are the sum of NM and DM masses. \label{fig:massrhoc} }
\end{figure}
\begin{figure}[ht!]
	\centering
	\includegraphics[width=1.0\linewidth]{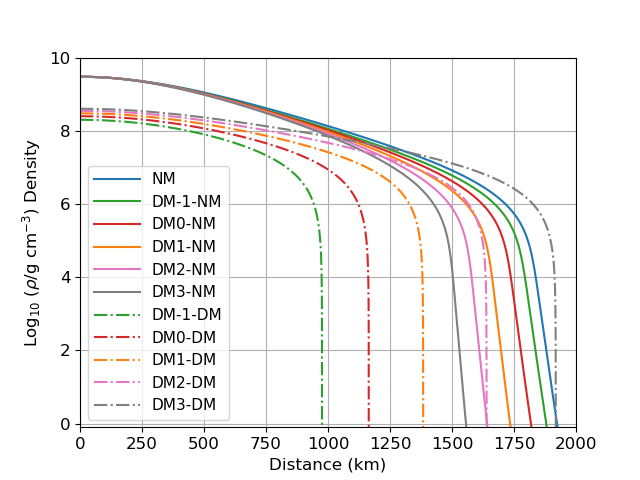}
	\caption{Density profiles for DM-admixed supernova progenitors. The solid (dashed) lines represent the NM (DM) densities for different amount of DM admixtures (see Table \ref{tab:starpara}). The solid blue line is for a pure NM WD. \label{fig:dmnmrho}}
\end{figure}
\begin{table}[!htb]
	\caption{Stellar parameters for different models of DM-admixed supernova progenitors. Ratios are computed as DM over NM. The DM particle mass is fixed at $0.1$ GeV. \label{tab:starpara}}
	\setlength\tabcolsep{0pt} 
	\footnotesize\centering
	\smallskip
	\begin{tabular*}{\columnwidth}{@{\extracolsep{\fill}}ccccccc}
		\hline \\ [0ex] 
		Model  & NM & DM-1 & DM0 & DM1 & DM2 & DM3 \\ [1ex] 
		\hline \\ [0ex] 
		DM $\rho_{c}$ ($10^{8}$ g cm$^{-3}$) & - & $2.0$ & $2.5$ & $3.0$ & $3.5$ & $4.0$ \\ [1ex]
		NM $\rho_{c}$ ($10^{9}$ g cm$^{-3}$) & $3.0$ & $3.0$ & $3.0$ & $3.0$ & $3.0$ & $3.0$ \\ [1ex]
		DM Mass ($M_{\odot}$) & - & $0.067$ & $0.120$ & $0.201$ & $0.322$ & $0.494$ \\ [1ex]
		NM Mass ($M_{\odot}$) & $1.374$ & $1.242$ & $1.183$ & $1.124$ & $1.067$ & $1.015$ \\ [1ex]
		DM Radius (km) & - & $975$ & $1160$ & $1380$ & $1640$ & $1920$ \\ [1ex]
		NM Radius (km) & $1930$ & $1890$ & $1830$ & $1740$ & $1650$ & $1560$ \\ [1ex]
	 	Radius Ratio & - & $0.52$ & $0.63$ & $0.79$ & $0.99$ & $1.23$ \\ [1ex]
		Mass Ratio & - & $0.05$ & $0.10$ & $0.18$ & $0.30$ & $0.47$ \\ [1ex]
		\hline
	\end{tabular*}
\end{table}
We use the Helmholtz EOS \citep{Timmes_2000} for NM assuming an initial constant temperature $T=10^{8}$ K. The initial composition of the NM is an equal mixture of $^{12}$C and $^{16}$O. We adopt the central density of NM to be $3\times10^{9}$ g cm$^{-3}$, which gives rise to thermonuclear explosion that is close to the average thermonuclear supernova \citep{Nomoto2017a, Kobayashi2020}. We vary the DM central density from $10^{8}$ g cm$^{-3}$ to $10^{10}$ g cm$^{-3}$ to obtain the masses and radii for the NM and DM components. We find that more DM admixture tends to decrease the NM mass which is similar to the results in previous literature using ideal cold EOS \citep{Leung_2013}. Figures \ref{fig:radrhoc} and \ref{fig:massrhoc} show that for lighter DM particle masses the DM radii are extended, and there could exist configurations where the DM and NM components are comparable in radii and masses. These results are in contrast with the result by \citet{Leung_2013} for heavier DM particle mass, where the DM form a compact point mass. Figure \ref{fig:massrhoc} shows that for a given NM mass, a vertical line that crosses the corresponding DM mass curve gives the minimum DM mass to be admixed so that the WD can develop towards the onset of explosion. For DM particle mass of $0.1$ GeV, there is a sharp transition of DM mass such that the DM mass increases drastically near the DM central density $8.75\times 10^{8}$ g cm$^{-3}$ while the NM mass changes relatively slowly. There are configurations where NM mass $\sim$1 $M_{\odot}$ while having nearly half as much of DM admixture.\\

We see that effects of the DM with particle mass $0.1$ GeV on the stellar parameters such as masses and radii are particularly interesting. We compute 5 configurations with such a DM particle mass as supernova progenitors, named as: DM-1, DM0, DM1, DM2, DM3. The model with no DM admixture is named as model NM. They are chosen in increasing ratio of DM to NM masses from $0.05$ to $\sim 0.5$. The parameters for the progenitor are shown in Table \ref{tab:starpara}, while the NM and DM density profiles of each model are shown in Figure \ref{fig:dmnmrho}. Model DM2 has nearly the same NM and DM radii, while model DM3 has an extended DM component that covers the NM completely. Note that in all models the NM is more massive than DM. We will investigate the thermonuclear explosions of these configurations. In particular, we are interested in how the extended DM component will affect the hydrodynamics and supernova observables, and how to distinguish normal supernovae from such DM-admixed supernovae.

\subsection{Explosion Hydrodynamics} \label{subsec:hydrodynamics}
\begin{figure}[ht!] 
	\centering
	\includegraphics[width=1.0\linewidth]{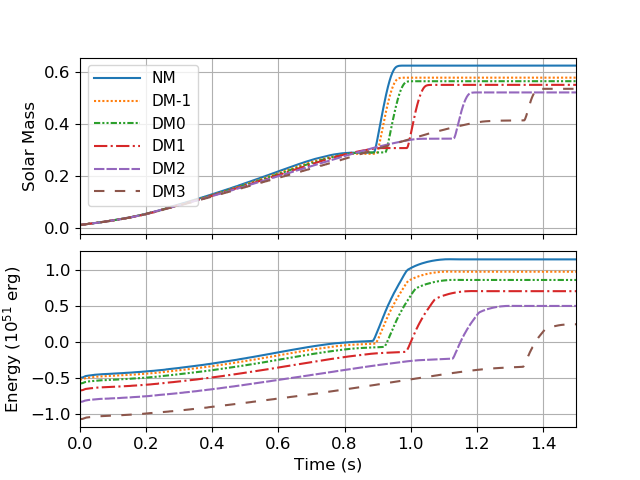}
	\caption{Total energy (bottom panel) and $^{56}$Ni mass (top panel) as a function of time for the DM-admixed models (DM-1, DM0, DM1, DM2, and DM3) compared to the pure NM one. See Table \ref{tab:starpara} for the description of the models. \label{fig:ni56energy}}
\end{figure}
\begin{figure}[ht!]
	\centering
	\includegraphics[width=1.0\linewidth]{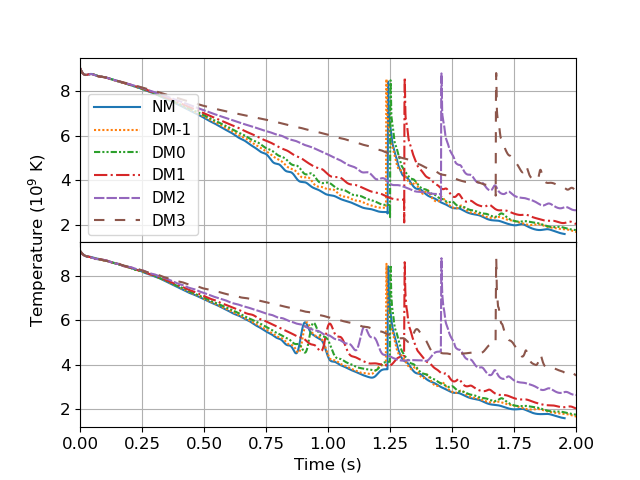}
	\caption{Same as Figure \ref{fig:ni56energy}, but for the maximum (bottom panel) and central temperature (top panel). \label{fig:tempcmax}}
\end{figure}
\begin{figure}[ht!]
	\centering
	\includegraphics[width=1.0\linewidth]{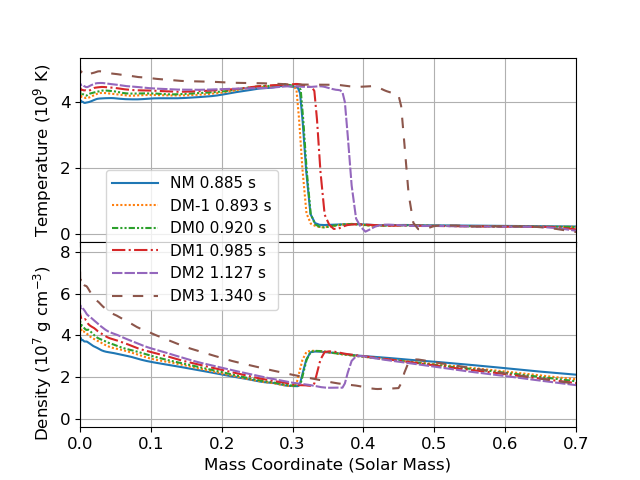}
	\caption{NM temperature (top panel) and density profiles (bottom panel) at the moment of DDT for different models. The number after each label gives the time of DDT in second. \label{fig:temprhoddt}}
\end{figure} 
\begin{figure}[ht!] 
	\centering
	\includegraphics[width=1.0\linewidth]{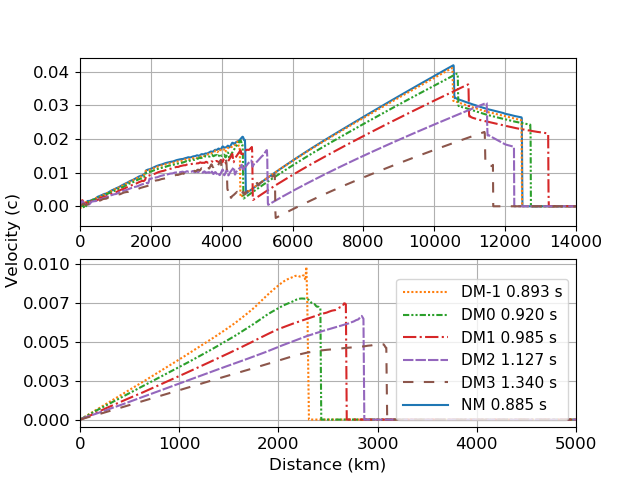}
	\caption{NM (top panel) and DM (bottom panel) velocity profiles at the moment of DDT for different models. Note the differences in the axis scales. \label{fig:dmnmvelddt}}
\end{figure} 
\begin{figure}[ht!] 
	\centering
	\includegraphics[width=1.0\linewidth]{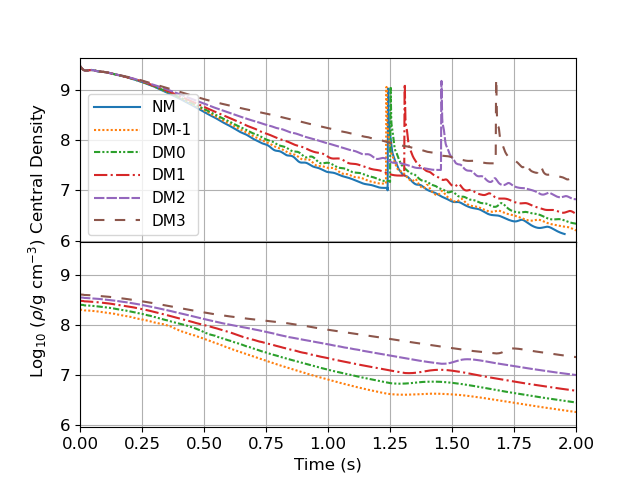}
	\caption{Same as Figure \ref{fig:ni56energy} but for DM (bottom panel) and NM (top panel) central densities. \label{fig:dmnmrhoc}}
\end{figure}
\begin{figure}[ht!] 
	\centering
	\includegraphics[width=1.0\linewidth]{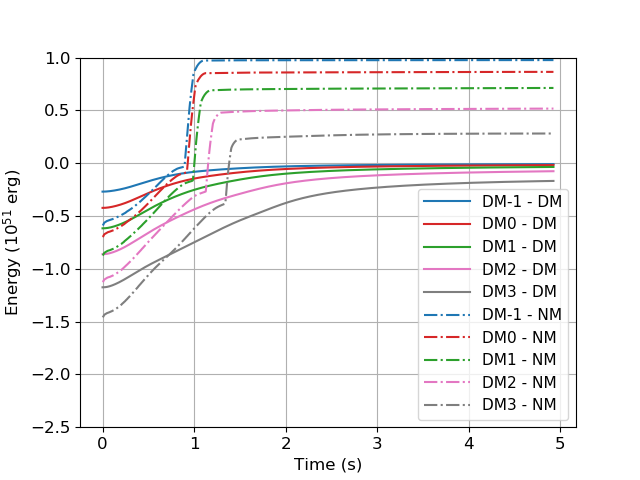}
	\caption{Same as Figure \ref{fig:ni56energy} but for DM (solid lines) and NM (dot-dashed lines) total energy. \label{fig:dmnmenergy}}
\end{figure} 
\begin{table}[!htb]
	\caption{Simulation results for models NM to DM3. $t_{\rm DDT}$ is the time of the transition to detonation. The last two rows represent the NM and DM energies at the end of the simulations.}
	\setlength\tabcolsep{0pt} 
	\footnotesize\centering
	\smallskip 
	\label{tab:simresults} 
	\begin{tabular*}{\columnwidth}{@{\extracolsep{\fill}}ccccccc}
		\hline \\ [0ex] 
		Model  & NM & DM-1 & DM0 & DM1 & DM2 & DM3 \\ [1ex] 
		\hline \\ [0ex] 
		$t_{\rm DDT}$ (s) & $0.885$ & $0.893$ & $0.920$ & $0.985$ & $1.127$ & $1.340$ \\ [1ex]
		$^{56}$Ni Mass ($M_{\odot}$) & $0.623$ & $0.577$ & $0.563$ & $0.549$ & $0.520$ & $0.534$ \\ [1ex]
		Energy ($10^{51}$ erg) & $1.650$ & $1.503$ & $1.442$ & $1.383$ & $1.336$ & $1.321$ \\ [1ex]
		NM Energy ($10^{51}$ erg) & $1.512$ & $0.977$ & $0.865$ & $0.713$ & $0.517$ & $0.281$ \\ [1ex]
		DM Energy ($10^{50}$ erg) & - & $-0.091$ & $-0.178$ & $-0.355$ & $-0.758$ & $-1.680$ \\ [1ex]
		\hline
	\end{tabular*}
\end{table}
The simulation results are summarized in Table \ref{tab:simresults}, where we list the energy, $^{56}$Ni generation and the time of transition to detonation. In general the increase in DM admixture delays the transition time and reduces the energy production. The DM gravitational potential and the interactions between DM and NM prohibit the expansion of NM during the deflagration phase which requires the NM to take a longer time to reach a lower density. On the other hand, the $^{56}$Ni masses synthesized are similar in magnitude. The $^{56}$Ni mass first decreases with more DM admixture to a minimum for model DM2, then rebounds proceeding to model DM3. We plot the total $^{56}$Ni mass variations in Figure \ref{fig:ni56energy}. In general, models with more DM admixture give lower $^{56}$Ni production after DDT. The steep density gradient reduces the mass content in the outer region of the NM, which is compensated by the higher production of $^{56}$Ni before DDT. The prolonged expansion during the deflagration phase gives more time to synthesize $^{56}$Ni. We also include the total energy variations in Figure \ref{fig:ni56energy}. It can be seen that the energy production after DDT is smaller with more DM admixture, which is consistent with the $^{56}$Ni variations. \\

We show the maximum and central temperatures in Figure \ref{fig:tempcmax}. We find that more DM admixture tends to keep the NM hotter. This can be explained by the suppressed expansion during the deflagration phase and thus NM cannot dissipate its internal energy by the pressure work done. Energy conservation suggests that the loss in internal energy would be converted not only to kinetic energy but also the gravitational energy. Hence for models with more DM admixture, the expansion velocity of NM is lower, as seen in Figures \ref{fig:temprhoddt} and \ref{fig:dmnmvelddt}, where the NM temperature, density, and velocity profiles of all models at the moment of DDT are shown. This also explains why these models have a later DDT transition time. The first peak in the maximum temperature corresponds to the transition to detonation, and it is consistent with the transition time $t_{\rm DDT}$ we recorded in Table \ref{tab:simresults}. \\

We plot the central densities of DM and NM in Figure \ref{fig:dmnmrhoc}. DM admixture tends to make NM expand slower, which is similar to the pattern found for temperature. The sudden increases in DM and NM central densities correspond to the second peaks of the central and maximum temperatures. This is because a converging acoustic wave generated during the DDT travels toward the center of NM and compresses and heats up the material. \\ 

We also tabulate the DM and NM energies in Table \ref{tab:simresults} and show their variations in Figure \ref{fig:dmnmenergy}. The energy $E_{i}$ for the $i^{\rm th}$ component is defined by 
\begin{equation}
	E_{i} = \sum_{\rm All Grids} \rho_{i}(\epsilon_{i}+ \phi_{j} + \frac{1}{2}(v_{i}^{2}+\phi_{i})).
\end{equation}
where $j \neq i$. The extended DM component has negative energy at the end for all of our considered models, which means that they remain bounded. Figure \ref{fig:dmnmenergy} shows that the NM and DM energies eventually approach constants towards the end of simulations. The increase in the DM energy after major exothermic nuclear reactions have finished is due to the fact that the contribution from NM gravitational potential is weaker when NM is expanding quickly. It hints at the decoupling between some of the most energetic NM and all bounded DM. The most energetic NM are governed by the explosion time scale which is much shorter than the dynamical time scale of DM. They can no longer transfer energy to unbind the remaining DM through gravitational interaction.

\subsection{Supernova Observables} \label{subsec:observed}
\begin{figure}[ht!] 
	\centering
	\includegraphics[width=1.0\linewidth]{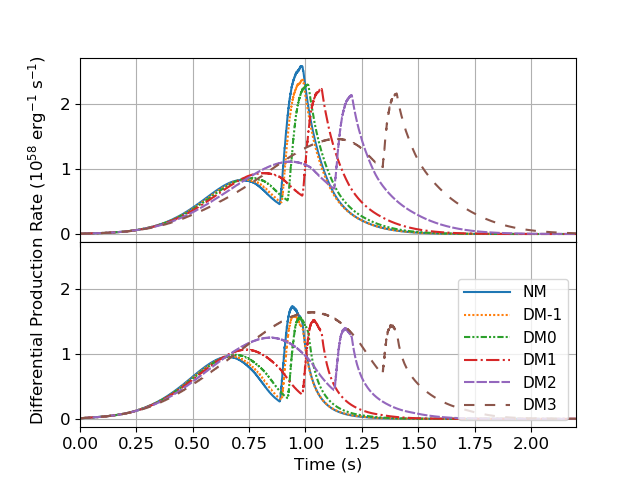}
	\caption{1 MeV (top panel) and 2 MeV (bottom panel) pair-neutrino differential production rates for different DM-admixed models (DM-1, DM0, DM1, DM2, DM3) compared to those of the pure NM model (solid blue line). \label{fig:1mev2mevpair}}
\end{figure} 
\begin{figure}[ht!]
	\centering
	\includegraphics[width=1.0\linewidth]{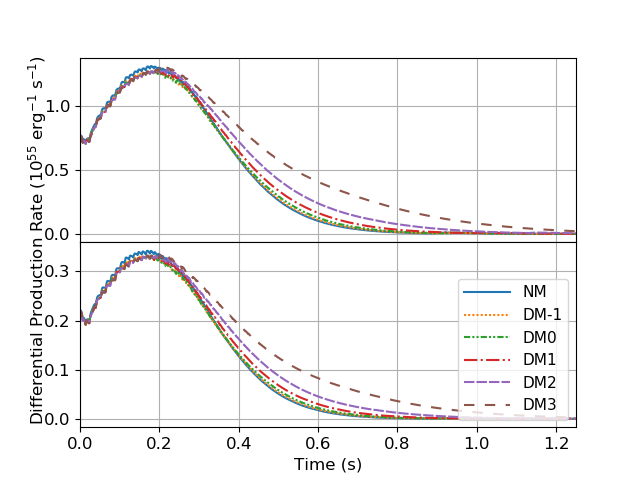}
	\caption{Same as Figure \ref{fig:1mev2mevpair}, but for the plasma-neutrino differential production rate. \label{fig:1mev2mevplas}}
\end{figure}
\begin{figure}[ht!] 
	\centering
	\includegraphics[width=1.0\linewidth]{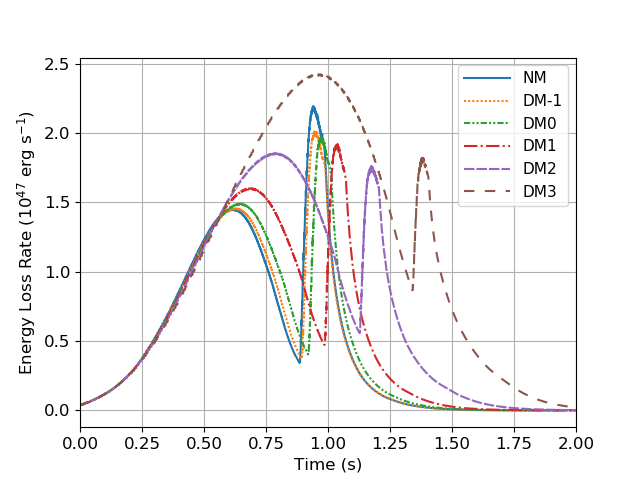}
	\caption{Same as Figure \ref{fig:1mev2mevpair}, but for the total thermo-neutrino energy loss. \label{fig:neuloss}}
\end{figure} 
\begin{figure}[ht!] 
	\centering
	\includegraphics[width=1.0\linewidth]{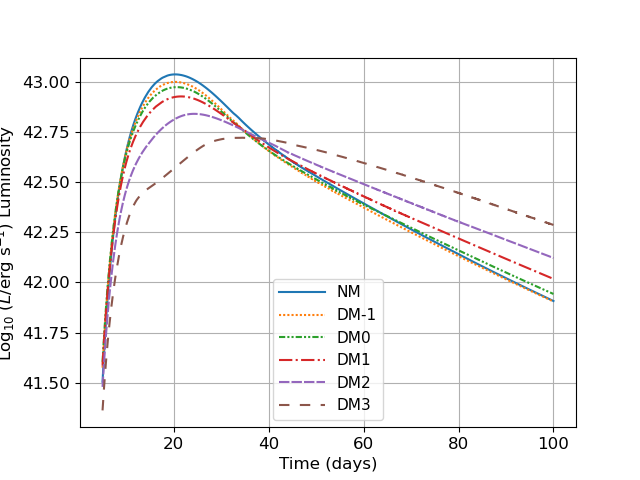}
	\caption{Light luminosity $L$ for different DM-admixed models (DM-1, DM0, DM1, DM2, DM3) compared to that of the pure NM model. (solid blue line) \label{fig:lumneu}}
\end{figure} 
\begin{figure} 
	\centering
	\includegraphics[width=1.0\linewidth]{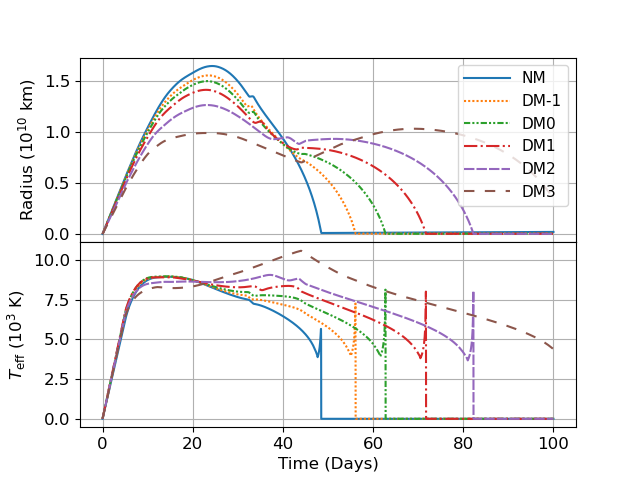}
	\caption{Same as Figure \ref{fig:1mev2mevpair}, but for the effective temperature $T_{\rm eff}$ (bottom panel) and radius of the photosphere (top panel). \label{fig:tempradphoto}}
\end{figure} 
\begin{table}[!htb]
	\caption{Neutrino time-integrated energy loss and production numbers for each MeV bin of neutrino energy from 1 -- 5 MeV, the sum over all bins (`Total'), and the ratio of the total neutrino numbers to that of the pure NM model.}
	\setlength\tabcolsep{0pt} 
	\footnotesize\centering
	\smallskip 
	\label{tab:neuobservabe} 
	\begin{tabular*}{\columnwidth}{@{\extracolsep{\fill}}ccccccc}
		\hline \\ [0ex] 
		Model  & NM & DM-1 & DM0 & DM1 & DM2 & DM3 \\ [1ex] 
		\hline \\ [0ex] 
		Energy ($10^{47}$ erg) & $0.970$ & $0.958$ & $0.998$ & $1.234$ & $1.426$ & $2.133$ \\ [1ex]
		1 MeV ($10^{52}$) & $1.232$ & $1.176$ & $1.204$ & $1.316$ & $1.595$ & $2.191$ \\ [1ex]
		2 MeV ($10^{52}$) & $1.944$ & $1.911$ & $1.993$ & $2.253$ & $2.875$ & $4.303$ \\ [1ex]
		3 MeV ($10^{52}$) & $1.264$ & $1.265$ & $1.329$ & $1.521$ & $1.981$ & $3.076$ \\ [1ex]
		4 MeV ($10^{52}$) & $0.605$ & $0.612$ & $0.643$ & $0.737$ & $0.959$ & $1.505$ \\ [1ex]
		5 MeV ($10^{52}$) & $0.249$ & $0.254$ & $0.266$ & $0.302$ & $0.389$ & $0.607$ \\ [1ex]
		Total ($10^{52}$) & $5.294$ & $5.218$ & $5.434$ & $6.129$ & $7.800$ & $11.682$ \\ [1ex]
		Ratio & $1.000$ & $0.986$ & $1.027$ & $1.158$ & $1.473$ & $2.207$ \\ [1ex]
		\hline
	\end{tabular*}
\end{table}
We tabulate the time-integrated total thermo-neutrino production and neutrino energy loss in Table \ref{tab:neuobservabe}. We find that models with more DM admixture tend to have higher neutrino energy loss and production in all MeV channels. We plot the neutrino energy loss rate and differential production rate at $1$ MeV and $2$ MeV in Figures \ref{fig:1mev2mevpair}, \ref{fig:1mev2mevplas} and \ref{fig:neuloss}. The sudden peaks in all neutrino observables at the transition time $t_{\rm DDT}$ are called `neutrino bursts' which have been observed in previous works \citep{PhysRevD.94.025026,PhysRevD.95.043006}. The neutrino burst signals are weaker for models with DM admixture when compared with the pure NM model, but the neutrino production for DM-admixed models are compensated by the prolonged deflagration phase, which is similar to what we find for $^{56}$Ni. We sum up the contributions of 1 -- 5 MeV thermo-neutrinos and compute the ratio of total neutrino production to that of the pure NM model. The total neutrino production for DM-admixed models can increase by a factor of $2.2$ when compared with the pure NM model. \\

There is an exception to the above discussion, which is model DM-1. This model has weaker neutrino observables when comparing with the pure NM model. As we discussed above the DM gravitational potential traps the NM, which suppresses the thermal expansion of the hot NM. However, the DM mass in DM-1 is only $0.067$ $M_{\odot}$. The gravitational potential is not deep enough to keep NM very hot when compared with the pure NM model (see Figure \ref{fig:temprhoddt}). On the other hand, the initial NM mass is substantially decreased by $0.132$ $M_{\odot}$. As a result, there is less NM to contribute to the production of thermo-neutrinos with similar temperature, and hence the neutrino signal is reduced. \\

After we terminate the simulation, we map the isotope, density, internal energy, electron fractions and velocity profiles into the SNEC code with uniform mass-coordinate. Since the hydrodynamic simulation shows that the DM remain bounded while the most energetic outer NM component is decoupled and ejected, the NM has a typical size much larger than that of the DM at the late phase of explosion. To approximate the gravitational effect from DM, we include the DM gravitational field as a point mass $g = -GM_{\rm DM}/r^{2}$ into the momentum equation of the SNEC code for all DM-admixed models. \\

The luminosity, radius and temperature of the photosphere are plotted in Figures \ref{fig:lumneu} and \ref{fig:tempradphoto}. We find that DM-admixed models give similar peak luminosities when compared with the pure NM model. In particular, the peak luminosity of model DM3 is only $2$ times lower than that of model NM. This can be understood as the amount of $^{56}$Ni generated are similar. The peak luminosity decreases for more DM-admixed models even though DM3 has slightly more $^{56}$Ni mass than DM2. For models with more DM admixture, the rise times are longer and the decline rates are slower than those of the pure NM model, producing a flatter and broader light curve. Note that in general, there exists a point of inflexion for a normal thermonuclear supernova light curve. We find that such a point disappears in the range of 0 -- 100 days for DM-admixed models with massive DM component. These features could be seen clearly for models DM2 and DM3. \\

Models with DM admixture tend to give higher luminosity in the early nebula phase (after 60 days, see \citet{10.1111/j.1365-2966.2007.12647.x} for details). The higher luminosity in the early nebula phase for DM-admixed models can be explained by the radii of photosphere. For the pure NM model, the photosphere shrinks to the center of the ejecta at around 50 days. The radii of photosphere for all DM-admixed models reach the center at a later time. The radius of photosphere even increases to a second peak for models DM2 and DM3, giving a larger effective temperature for the photosphere at the early nebula phase.

\section{Discussion} \label{sec:discussion} 
\subsection{Effect Of Extended DM Component} \label{subsec:dmcomponent}
A DMWD with DM particle mass of $0.1$ GeV has interesting equilibrium structure. The DM component no longer acts as a point mass but has an extended radius, which is comparable to that of NM. Whether the subsequent supernova is different from that of a pure NM model depends on the competition between the NM and DM contents. In our considered models, the NM mass content does not change significantly when compared with the DM (see for example, Table \ref{tab:starpara} and Figure \ref{fig:dmnmrho}). We show that the extra DM-NM gravitational interaction would have significant impact on the expansion phase of deflagration, keeping matter hot and dense and producing more $^{56}$Ni during that period. In particular, with model DM3 which has less NM content than DM2, we find more $^{56}$Ni being produced. What if the DM content increases even further? Would it be possible to synthesize more $^{56}$Ni than a pure NM model? The DM content may become too massive such that no NM could escape. It would then behave like a failed detonation \citep{Kasen_2007, Plewa_2007, Jordan_2012} even though DDT occurs. It's an interesting future work to model the thermonuclear flame reignition and nuclear reactions of such a model.

\subsection{Formation Of Compact Dark Star} \label{subsec:darkstar}
We show that the DM are left behind as a compact dark star with mass ranging from $\sim 0.07$ to $0.5$ $M_{\odot}$, in all of our considered models. This suggests an alternative way to search for thermonuclear supernova diversity and hence astrophysical dark matter - to look for any dark compact remnant in thermonuclear supernova events through for example, micro-lensing effect. We note that recent development in gravitational-wave astronomy has opened-up a new window to search for astronomical compact objects, especially sub-solar-mass black holes \citep{Shandera_2018, alex2020search}. Since dark stars may mimic black holes as dark gravitating sources, our results show that if there are observed sub-solar-mass dark gravitational sources, they could be compact dark stars remaining after a DM-admixed supernova explosion with DM admixture.\footnote{However, one distinctive feature to distinguish between a dark star and a black hole will be the absence of the event horizon in the former one.}

\subsection{Neutrino Signal} \label{subsec:neutrino}
\begin{table}[!htb]
	\caption{Neutrino event estimations for different models. The first column lists 4 different detectors. For the LENA detectors two different detection methods are presented, one using elastic scattering off electrons (ES0) and the other elastic scattering off protons (PES). The neutrino event rate calculations for different detectors are based on the work of \citet{supernovaneutrino}. No neutrino threshold energy is assumed in our rough estimations. }
	\setlength\tabcolsep{0pt} 
	\footnotesize\centering
	\smallskip 
	\label{tab:neutable} 
	\begin{tabular*}{\columnwidth}{@{\extracolsep{\fill}}ccccccc}
		\hline \\ [0ex] 
		-  & NM & DM-1 & DM0 & DM1 & DM2 & DM3 \\ [1ex] 
		\hline \\ [0ex] 
		Hyper-K, Memphys & $3.3$ & $3.3$ & $3.4$ & $3.8$ & $4.9$ & $7.3$ \\ [1ex]
		Glacier & $4.2$ & $4.1$ & $4.3$ & $4.9$ & $6.2$ & $9.3$ \\ [1ex]
		LENA (ES0) & $2.9$ & $2.9$ & $3.0$ & $3.4$ & $4.3$ & $6.4$ \\ [1ex]
		LENA (PES) & $12.0$ & $11.8$ & $12.3$ & $13.9$ & $17.7$ & $26.5$ \\ [1ex]
		\hline
	\end{tabular*}
\end{table}
Models with more DM admixture tend to produce more thermo-neutrino signals. It would be interesting to estimate the number of expected neutrino events for the models considered in this work. It was shown that the number of events $N$ scale as \citep{doi:10.1080/0142241042000198877}:\\
\begin{equation}\label{scale}
	N = N_{t} \times \sigma_{\nu} \times \frac{\dot{N}_{\nu}}{4\pi d^{2}} \times \tau,
\end{equation}
where $N_{t}$ is the number of targets, $\sigma_{\nu}$ is the neutrino interaction cross section, $\dot{N}_{\nu}$ is the number of neutrinos produced per second, $d$ is the distance of the supernova from Earth and $\tau$ is the time elapsed. We notice that there are previous results estimating the event counts for some detectors proposed or under construction \citep{supernovaneutrino}. They include Hyper-Kamiokande \citep{hyperkamiok2018hyperkamiokande}, Memphys \citep{Autiero_2007,Patzak_2012}, Glacier \citep{Autiero_2007,Rubbia_2009} and LENA \citep{Autiero_2007,WURM2015376}. Here we use Equation (\ref{scale}) to estimate the ratios in Table \ref{tab:neuobservabe}. The results are presented in Table \ref{tab:neutable}. In general, since the thermo-neutrino signal is weak, the difference in neutrino production is distinguishable only for models DM2 and DM3. We have not done post-processing on weak interaction neutrinos due to the limited number of isotopes. This together with a more detailed analysis in nucleosynthesis using a full nuclear network will be an interesting research direction in the future.

\subsection{Effect Of DM Gravity} \label{subsec:dmgravity}
\begin{figure}[ht!]  
	\centering
	\includegraphics[width=1.0\linewidth]{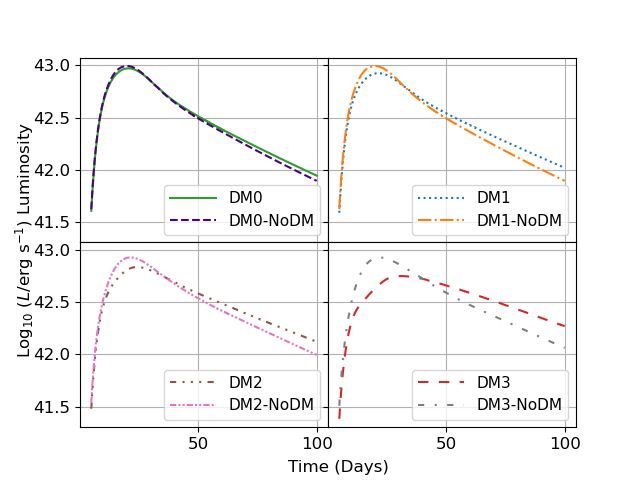}
	\caption{Luminosities with and without (labeled NoDM) including the DM gravity in the momentum equation of the SNEC code. \label{fig:lumnodm}}
\end{figure}
Although the amount of $^{56}$Ni generated for model DM3 is more than that of DM2, the peak luminosity is lower. Moreover, the light curves are flatter and broader, and their points of inflexion disappear as more DM are admixed. It is interesting to study which factor governs the peak luminosity and shape of the light curve. In particular, the major difference between DM2 and DM3 is the amount of DM remnant. To explore the effect of DM gravity, we add one more light curve for each of the models DM0 -- DM3 in Figure \ref{fig:lumnodm}. They are labelled as `NoDM' and they do not include the DM gravity in the momentum equation of the SNEC code. We find that the presence of DM gravitational force significantly alters the peak luminosity as well as the shape of the light curve for models with massive DM component - the light curves become narrower, and their points of inflexion re-appear, especially for models DM2 and DM3, after we switch off the DM gravity. Thus, we conclude that the effects of DM cannot be neglected even in the light curve calculation.

\subsection{Is there any trapped NM?} \label{subsec:trappednm}
We note that although the NM in all models have positive energies at the end of the simulation, it is possible that there is a small but non-zero amount of bounded NM at the tail of the ejecta, especially in the zone that is overlapping with DM. Nevertheless, we are ultimately interested in the supernova observables and the amount of unbounded NM is important only in calculating the light curve using analytic formulae \citet{Dado_2015}. In our studies, we have already added the effect from DM consistently in the SNEC's momentum equation, and so the SNEC code would keep track of the falling NM, which makes estimating the bounded NM mass not necessary. Moreover, the fate of the NM in the post-explosion phase is governed by the $^{56}$Ni and $^{56}$Co decay and $\gamma$-ray deposition. The kinetic energy transfer from photons would also help increase the kinetic energy for NM to escape the DM potential. In short, to model the radiative transfer of such a model will require a more realistic two-fluid radiative transfer code, involving complex gamma-ray heating in the bounded core which is beyond the scope of this study.

\subsection{Relation To Peculiar Thermonuclear Supernovae} \label{subsec:slow}
\begin{figure}[ht!] 
	\centering
	\includegraphics[width=1.0\linewidth]{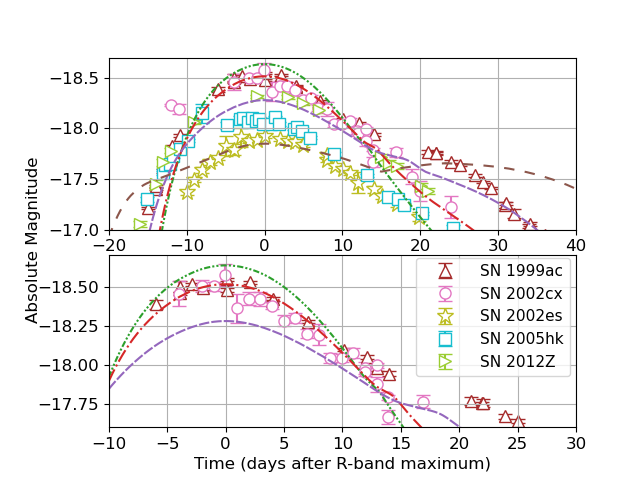}
	\caption{Comparison of R-band light curves of our models with those of the observed peculiar thermonuclear supernova events. Data are extracted from The Open Supernova Catalog \citep{2017ApJ...835...64G}. Note that we have artificially off set the light curves in the temporal direction, in order to match with the light curves of our models. See Figure \ref{fig:lumneu} for description of our models’ light curves. The lower panel is the zoom in plots for models DM0, DM1 and DM2.  \label{fig:lcscatter}}
\end{figure}
\begin{figure}[ht!] 
	\centering
	\includegraphics[width=1.0\linewidth]{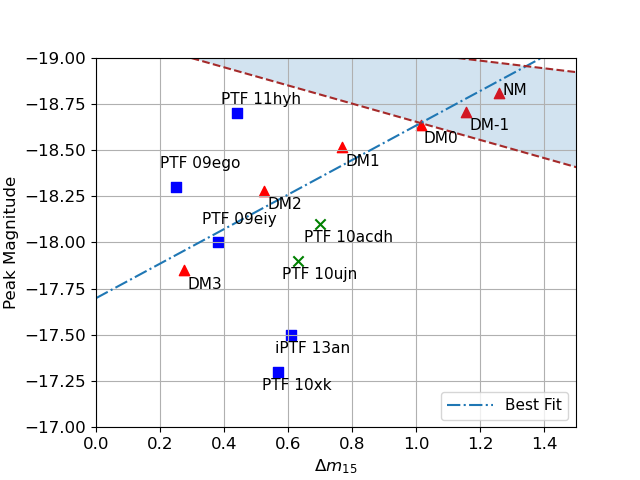}
	\caption{The light curve peak magnitudes against the decline rate $\Delta m_{15}$ for our models and
		some observed events taken from \citet{White_2015}. We divide the observed events according to treatment of the original text. The blue marked events are SN 2002cx-like, while the green ones are SN 2002es-like. Error bars are not shown since they are not given in the original text. The best fit line gives a slope of $-0.934$ and intercept $-17.698$. The shaded area is spanned by the R-Band Phillip's relation computed by \citet{Prieto_2006}, using supernova samples from \citet{Jha_2007}. \label{fig:peaklumscatter}}
\end{figure}
Peculiar thermonuclear supernovae had been observed in the past few decades. Depending on their luminosity and spectra, they can be further classified as super-luminous \citep[e.g. SN 1991T][]{1992AJ....103.1632P}, sub-luminous \citep[e.g. SN 1991bg][]{1992AJ....104.1543F}, and sub-luminous with strong mixing (also known as Type Iax supernovae). Among many Type Iax supernovae observed, SN 2002cx is one of the classical example \citep{Li_2003}. In particular, some of these events show a slower decline rate than those of normal supernovae. We try to explain some of these events by DM-admixed thermonuclear supernovae. We extracted R-band light curves\footnote{We follow the treatment by \citet{White_2015} to consider only R-band magnitude, so as to be consistent with the discussion in the next section.} of the following supernova events: SN 1999ac \citep{Candia_2003,Ganeshalingam_2010,Silverman_2012}, SN 2002cx, SN 2002es \citep{Ganeshalingam_2012}, SN 2005hk \citep{Chornock_2006,Phillips_2007,Silverman_2012,Stahl2019}, SN 2012Z \citep{Silverman_2012,Stahl2019}, from The Open Supernova Catalog \citep{White_2015}, to be shown together with the R-band light curves of our models in Figure \ref{fig:lcscatter}. We find that SN 1999ac and SN 2002cx are well fitted by the DM1 model before 15 days post R-band maximum\footnote{In general a multi-frequency radiative transfer code is necessary for a consistent prediction of band-specific luminosity. In this work, we use the embedded R-band filter implemented in SNEC to isolate the R-band luminosity based on the black-body distribution. We only consider the period where the photosphere is hot and dense, so that local thermodynamical equilibrium approximation can be valid. The frequency-dependent opacity, for example, will be necessary in order to capture the later evolution, for example, the second maximum in R-band light curves.}, suggesting these events could be possible for DM-admixed thermonuclear supernovae.\\

Recent studies using observational data from Palomar Transient Factory \citep[PTF,][]{Law_2009} also found peculiar thermonuclear supernovae having dimmer luminosities, longer rise times and slower decline rates \citep{White_2015}. Such features resemble the drop of peak luminosity and flattened light curve in our models. We plot the light curves' peak magnitudes (R-band)\footnote{Model provided by that discovery paper provides light curves data on R-band only.} against the decline rate $\Delta m_{15}$, for the observed supernova events through PTF: PTF 09ego, PTF 09eiy, PTF 10xk, PTF 11hyh, iPTF 13an, PTF 10ujn and PTF 10acdh, together with those of our models in Figure \ref{fig:peaklumscatter}. Our models show a wide range of peak magnitudes ranging from $-17.8$ to $-18.8$, and decline rates from $0.28$ to $1.26$. We find that PTF 09eiy data are closest to the best-fit lines for the series of DM-admixed models in Figure \ref{fig:peaklumscatter}, followed by PTF 09ego, PTF 10acdh and PTF 10ujn. These events seem to be possible candidates for DM-admixed thermonuclear supernovae. \\

We show that the less luminous light curves give slower decline rates, which breaks Phillip's relation \citep{Phillips_1999} for Type Ia supernova light curves. Our results suggest that the DM admixture provides the extra degree of freedom to explain the Type Ia supernova diversity, based on its orthogonality trend to the Phillips' relation. The diversity of Type Ia supernova along the Phillip's relation is likely to be associated with the change of canonical model parameters, such as mass, metallicity and explosion kernels. Besides other explosion mechanism, the DM admixture provides an alternative to explain the Type Ia supernovae apart from the band. 

\subsection{Limitations And Improvements} \label{subsec:short}
Beside a simple 1D modelling for the explosion scenario, several limitations of this work should be noted. We constructed the series of Type Ia supernova models based on a model without DM. As discussed in \cite{Leung_2018}, to match with the diversified Type Ia supernova data, a wide range of `normal' models, coupled with different amounts of DM admixture, is necessary to span the parameter space. Given that the effects of DM observed in this work is generic, we expect that the trend when DM admixed with other ordinary Type Ia supernova models, will be similar. \\

Another caveat is that for Type Iax supernovae, their spectra show strong mixing features and weak intermediate mass elements lines. Our DM-admixed models show that the delayed DDT can strongly suppress the strength of detonation. This suggests that lower amounts of intermediate mass elements are produced. However, a concrete conclusion will require combination of radiative transfer with an extended network of nuclear reactions to capture the explosive nucleosynthesis, which will be an interesting future project. \\

Last but not least, we note that our calculation does not include full radiation transport, and the spectrum of different light frequencies is omitted. It would be interesting to investigate the post-explosion light curves using multi-band radiative transfer models. We have shown the importance of the DM gravity to the observed light curve. Including the DM gravitational interaction terms in the radiative transfer phase is necessary for a more accurate treatment. 

\section{Conclusion} \label{sec:conclusion}
We present delayed-detonation simulations of thermonuclear supernovae admixed with extended near-solar-mass DM. DM admixture plays an important role in the expansion phase by deflagration. DM-admixed models tend to give out stronger neutrino signals with comparable iron-group elements to those of ordinary models without DM admixture. Our numerical models show dimmer and broader light curves, which challenge Type Ia supernovae as standard candles in distance measurement. We propose some observed supernova events as possible candidates of DM-admixed thermonuclear supernovae. Our results also show that the DM component will remain bounded as a compact remnant which mimics sub-solar-mass black holes as a dark gravitational source.

\acknowledgments
We thank K.-W. Wong for his development of a two-fluid hydrodynamics solver upon which our code is constructed. We also thank F. X. Timmes for his open-source codes of the Helmholtz EOS and thermo-neutrino emission package. S.-C. Leung acknowledges support by funding HST-AR-15021.001-A and 80NSSC18K1017. This work is partially supported by a grant from the Research Grant Council of the Hong Kong Special Administrative Region, China (Project No. 14300320). 

\software{SNEC: \citet{Morozova_2015}}

\appendix
\section{Effect Of A Movable DM Component} \label{sec:movabledm}
Extended from a previous work by \citet{Leung_2015} where the DM is assumed to be a stationary point mass, this work takes into account the DM dynamics. Here we explain why, even when the DM couples to the NM only by gravity, the DM motion is still important for the deflagration, where the DM has a comparable size as the NM. To illustrate the effects, we repeat Models DM-1 and DM3, but with the DM motion frozen. These models are named: DM-1-Static, DM3-Static. These two models stand for two extremes in how the DM affects the WD. Model DM-1 approaching the limit in the work by \citet{Leung_2015}, where more DM admixture can reduce $^{56}$Ni production in the PTD explosion. The simulation results are presented in Table \ref{tab:movabledmtable}. Our result shows that freezing the DM motion tends to under produce the crucial supernova observables. Hence not only the DM gravity, but also the dynamical interaction between DM and NM through gravity are important in our studies.
\begin{table}[!htb]
	\begin{center}
	\caption{Simulations results for frozen DM-admixed models DM-1 and DM3 comparing with their movable counterparts.}
\setlength\tabcolsep{0pt} 
\footnotesize\centering
\smallskip 
\label{tab:movabledmtable} 
\begin{tabular*}{\columnwidth}{@{\extracolsep{\fill}}ccccc}
	\hline \\ [0ex] 
	Model  & DM-1 & DM-1-Static & DM3 & DM3-Static \\ [1ex] 
	\hline \\ [0ex] 
	$t_{\rm DDT}$ (s) & $0.893$ & $0.639$ & $1.340$ & $0.404$ \\ [1ex]
 	$^{56}$Ni ($M_{\odot}$) & $0.577$ & $0.430$ & $0.534$ & $0.232$ \\ [1ex]
	Neutrino Loss ($10^{47}$ erg) & $0.958$ & $0.499$ & $2.133$ & $2.037$ \\ [1ex]
	Total Neutrino Production ($10^{52}$) & $5.218$ & $2.784$ & $11.682$ & $1.141$ \\ [1ex]
	\hline
\end{tabular*}
\end{center}
\end{table}

\section{Formation Of DMWD} \label{sec:dmawdcreate}
It was pointed out by \citet{Iorio_2010} that a neutron star in the Galaxy could accrete DM at a rate as high as $10^{7}$ kg s$^{-1}$. However, at this constant rate, it will take $\sim 6.3\times10^{6}$ Gy to accrete DM to about $1$ $M_{\odot}$, which is much longer than the age of the universe \citep{planck2020}. We consider the scenario similar to that in \citet{Leung_2013} where the star is born with an inherent admixture of DM, where DM contributes the gravity since zero-age main-sequence star. By computing the Jean's radius of the molecular cloud collapse, the required density for DM to form a $0.01$ $M_{\odot}$ DM admixture is about $1$ GeV cm$^{-3}$. Here we do a rough estimation on the required density for our considered models. We take the most extreme model with DM mass $0.5$ $M_{\odot}$. Since the NM mass for a thermonuclear supernova progenitor is around $3$ to $7$ $M_{\odot}$, we have DM mass $\ll$ NM mass, and the Jean's radius should remain almost unchanged in order of magnitude. Since $\rho \propto M$, we have the required density as $50$ GeV cm$^{-3}$. It was suggested by \citet{SANDIN2009278} that the typical DM halo density ranges from tenth to hundreds of GeV cm$^{-3}$, which shows that it is possible for near-solar-mass DM admixture in WDs.

\bibliography{paper}{}
\bibliographystyle{aasjournal}

\end{document}